# On the mystery of the absence of a spin-orbit gap in scanning tunneling microscopy spectra of germanene

Carolien Castenmiller and Harold J.W. Zandvliet[&]

Physics of Interfaces and Nanomaterials & MESA+ Institute for Nanotechnology, University of Twente, P.O. Box 217, 7500AE Enschede, The Netherlands

**Abstract**

Germanene, the germanium analogue of graphene, shares many properties with its carbon counterpart. Both materials are two-dimensional materials that host Dirac fermions. There are, however, also a few important differences between these two materials: (1) graphene has a planar honeycomb lattice, whereas germanene's honeycomb lattice is buckled and (2) the spin-orbit gap in germanene is predicted to be about three orders of magnitude larger than the spin-orbit gap in graphene (24 meV for germanene versus 20 µeV for graphene). Surprisingly, scanning tunneling spectra recorded on germanene layers synthesized on different substrates do not show any sign of the presence of a spin-orbit gap in germanene. To date the exact origin of the absence of this spin-orbit gap in the scanning tunneling spectra of germanene has remained a mystery. In this work we show that the absence of the spin-orbit can be explained by germanene's exceptionally low work function of only 3.8 eV. The difference in work function between germanene and the scanning tunneling microscopy tip (the work functions of most commonly used STM tips are in the range of 4.5 to 5.5 eV) gives rise to an electric field in the tunnel junction. This electric field results in a strong suppression of the size of the spin-orbit gap.

[&] Corresponding author: h.j.w.zandvliet@utwente.nl

**Introduction**

Since the rise of graphene [1,2] there have been many attempts to grow or synthesize other two-dimensional materials that have properties that are similar or comparable to graphene. The low-energy electrons of graphene, i.e. the electrons in the vicinity of the Fermi level, have a linear energy-momentum dispersion relation. Owing to this linear dispersion relation these electrons behave as massless relativistic particles. The most appealing surrogates of graphene consist of the elements that can be found in the same column of the periodic system as carbon. Silicon, germanium and tin have an electronic configuration that is very similar to carbon. All these elements have four valence electrons, two electrons in an *s*-shell and two electrons in a *p*-shell. Already in the mid-90s of the previous century Takeda and Shiraishi pointed out in a theoretical study that the silicon and germanium analogues of graphite are in principle stable [3]. Takeda and Shiraishi showed that the only difference with graphite is that the two-dimensional honeycomb lattices of silicon and germanium are not flat, but buckled (see Figure 1). Despite this buckling the silicon and germanium analogues of graphene, which are referred to as silicene and germanene, respectively, are predicted to host Dirac fermions. Unfortunately, these materials do not occur in nature and therefore there have been many attempts to grow or synthesize these materials [4-6].

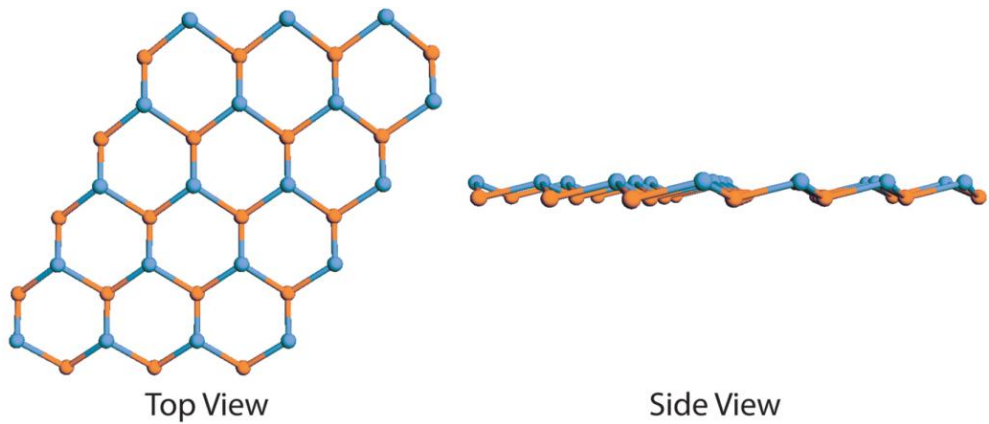

**Figure 1** *Buckled honeycomb lattice of germanene. Left: top view. Right: side view.*

Carbon occurs in nature in two allotropes: diamond and graphite. The diamond structure consists of *sp³* hybridized carbon atoms, whereas the carbon atoms in the graphite structure are *sp²* hybridized. At room temperature and atmospheric pressure the graphite structure has a lower energy than the diamond structure, which implies that diamonds are not forever. Unfortunately, diamond will eventually convert to graphite. In the case of silicon and germanium, however, this situation is reversed, i.e. the diamond structure is lower in energy than the layered graphite structure. Synthesizing or growing a single layer of silicene or germanene seems therefore, at least at first sight, a mission impossible. Fortunately, a single layer of $sp^3$ hybridized silicon or germanium turns out to be instable. Cahangirov et al. [7] theoretically studied two-dimensional silicon and germanium layers using density functional theory. They considered three different structures: (1) the planar honeycomb structure, (2) the low-buckled honeycomb $sp^2/sp^3$ hybridized structure and (3) the high-buckled $sp^3$ hybridized honeycomb structure. The last structure, the high-buckled configuration, has the lowest energy for both silicon and germanium. Cahangirov et al. [7] also calculated the phonon spectra of all these structures and found that the high-buckled configurations have imaginary phonon modes in a substantial part of the Brillouin zone. This means that the high-buckled configurations have to be discarded as they are not stable! The stable structure with the lowest energy is, for silicon as well as for germanium, the low-buckled configuration. The planar silicene and germanene honeycomb lattices are metallic, whereas the low-buckled lattices are semimetals, provided at least that spin-orbit coupling is not taken into account [8]. If spin-orbit coupling is considered a small bandgap opens at the K and K' points of the surface Brillouin zone.

In 2005 Kane and Mele [9,10] showed that graphene is a two-dimensional topological insulator that should in principle exhibit the quantum spin Hall effect. Owing to the spin-orbit coupling in graphene, the interior of the material is gapped, whereas the edges are gapless. The metallic states at the edges are spin-polarized and topologically protected. Unfortunately, the spin-orbit coupling in graphene is very weak resulting in a spin-orbit gap of only ~ 20 μeV. Therefore the quantum spin Hall effect in graphene is only observable at extremely low temperatures. Elements such as Si and Ge exhibit a much stronger spin-orbit coupling, since the spin-orbit coupling scales as $Z^4$, where $Z$ is the atomic number. In a theoretical study Liu, Feng and Yao [11] reported that the spin-orbit gaps in silicene and germanene are 1.55 meV. 23.9 meV, respectively. Particularly germanene seems to be an ideal candidate to exhibit the quantum spin Hall effect at moderate temperatures [8,11]. However, no evidence for the presence of a spin-orbit gap in scanning tunneling spectra of germanene has been found yet. The exact reason for the absence of this spin-orbit gap has remained a mystery.

In this paper we will scrutinize all the available scanning tunneling spectroscopy measurements that have been performed on germanene. We will elaborate on the effect that an electric field in the scanning tunneling microscopy junction can have on the size of the spin-orbit gap. We will show that the key to the absence of a spin-orbit gap in scanning tunneling spectra is the low work function of germanene. Germanene has a work function of only 3.8 eV [12], which is substantially lower than the work function of materials that are used for scanning tunneling microscopy tips. This difference in work function means that there is always an electric field present in the germanene-scanning tunneling microscopy tip junction. We will show that this electric field results in a decrease of the spin-orbit gap in germanene.

**Results and discussion**

In the literature there are three germanene systems for which scanning tunneling spectroscopy experiments are performed. We will discuss these three examples one-by-one and in chronological order. The first system is germanene synthesized on Ge$_2$Pt clusters [5]. The germanene forms upon cooling down eutectic Pt$_{0.22}$Ge$_{0.78}$ droplets, which undergo a spinodal decomposition into a Ge$_2$Pt alloy and a pure germanium phase at temperatures below the eutectic temperature (~1040 K) [13]. Subsequently, the excess germanium segregates to the surface of the Pt/Ge clusters because germanium has a lower surface free energy per unit area than Ge$_2$Pt. There is ample of evidence that the Ge$_2$Pt clusters are not coated with a single layer, but rather a few layers of germanium. Bampoulis *et al.* [5] showed that the outermost layer exhibits a buckled honeycomb structure with a lattice constant of about 4.2 Å. A year later, Zhang *et al.* [14] performed scanning tunneling spectroscopy measurements on the outermost germanene layer at room temperature. The scanning tunneling spectra revealed a well-defined V-shaped density of states, which is one of the hallmarks of a two-dimensional Dirac material [14]. The minimum of the V-shaped density of states is, however, a bit rounded. Walhout *et al.* [15] analyzed the exact shape of the V-shaped density of states at room temperature as well as at 77 K and found that in both cases the rounding can be explained by thermal broadening. These authors also arrived at the conclusion that if there is a spin-orbit gap in germanene, it has to be smaller than 6 meV. In the second series of experiments germanene was grown on molybdenum disulfide, a transition metal dichalcogenide with a bulk bandgap of ~1.3 eV. Room temperature scanning tunneling spectroscopy experiments revealed a V-shaped density of states without any sign of the presence of a spin-orbit gap [16]. The third series of experiments deals with the system germanene on Cu(111) [17]. Qin *et al.* [17] deposited germanium on Cu(111) and found that the first germanene layer is electronically coupled to the Cu(111) substrate and does not exhibit a V-shaped density of states. The second germanene layer, however, is decoupled from the Cu(111) substrate and displays a well-defined V-shaped density of states. Moreover, also this spectrum shows no sign whatsoever of the presence of a spin-orbit gap [17].

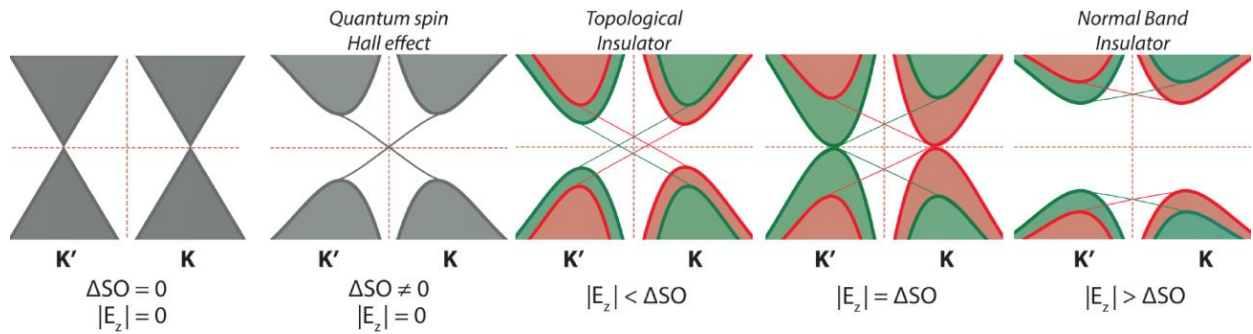

**Figure 2** *Schematic diagram of the band structure of germanene near the K and K' points of the surface Brillouin zone. From left to right: band structure without a spin-orbit gap, with a spin-orbit gap, applied electric field smaller than the critical value, applied electric field equal to the critical value and applied electric field larger than the critical value. Red and green refer to spin up and spin down bands, respectively.*

As we will argue below the origin for the absence or strongly suppressed spin-orbit gap of germanene in scanning tunneling spectroscopy is caused by the electric field, $E_z$, that is always present in the tunnel junction. The electric field in a scanning tunneling microscopy junction has in principle two contributions. The first contribution of the electric field originates from the bias that is applied across the tunnel junction. This results in an electric field with a strength $V/d$, where $V$ is the applied sample bias and $d$ the width of the tunneling gap (typically 1 nm). As the Dirac point of germanene is in all three examples very close to the Fermi level, this electric field component is rather small. The second contribution to the electric field is related to the difference in work function of germanene and the scanning tunneling microscopy tip. In a recent paper Borca *et al.* [12] measured the work function of germanene and germanium using field emission resonances by recording the derivate of the z-piezo displacement to the bias voltage, $dz/dV$, at sample bias voltages that exceed the work function of both materials. This method provides more reliable and accurate estimates for the work function than the standard $dI/dz$ scanning tunneling spectroscopy measurements, which is recorded at moderate voltages. Borca *et al.* [12] found a value of 3.8 eV and 4.5 eV for germanene on $Ge_2Pt$ and germanium, respectively. The measured work function of germanium, which served as a reference in these experiments, is in perfect agreement with available experimental data. As both spectra are recorded with the same scanning tunneling microscopy tip, this gives confidence that also the work function of germanene on $Ge_2Pt$ is correct. Although the work function of a material can be affected, albeit slightly, by the choice of substrate, a comparable value of the work function is expected for the other two germanene systems. The scanning tunneling spectra of germanene that have been discussed above were all recorded with a tungsten (W) scanning tunneling microscopy tip [5,13,14,17]. Tungsten has a work function of 4.5 eV, which is ~0.7 eV larger than the work function of germanene. This means that the electric field in the tunnel junction of germanene and scanning tunneling microscopy tip is ~0.7 eV/nm (the width of the tunnel gap is estimated to be about 1 nm). Due to the buckling of the honeycomb lattice, the external electric field causes a shift of charge from one triangular sublattice to the other triangular sublattice of germanene. This charge shift has a profound effect on the spin-orbit gap of the germanene. The energy dispersion in the vicinity of the *K* and *K'* points of the Brillouin zone of germanene is given by [18],

$$E_{\pm} = \pm\sqrt{\hbar^2 v_F^2 k^2 + \left(\frac{\Delta}{2}eE_z - \zeta s \lambda_{SO}\right)^2} \qquad (1)$$

where *ζ=±1* refers to the *K* (*K'*) point, *s=±1* to the spin, $\Delta$ to the buckling of germanene, *e* to the unit of elementary charge, $\lambda_{SO}$ to the spin-orbit gap, $v_F$ to the Fermi velocity, *k* to the momentum and $\hbar$ to the reduced Planck's constant. For a vanishing spin-orbit coupling and electric field the well-known Dirac cones are recovered, i.e. $E_{\pm} = \pm\hbar v_F k$. The $(eE_z - \zeta s \lambda_{SO})$ term makes that the energy-momentum dispersion is not perfectly linear anymore. As can be seen from Eq. (1) a small electric field results in a decrease of the bandgap, but the germanene remains a two-dimensional topological insulator (see Figure 2). The bandgap closes completely at a critical electric field, $E_{c,z}$, given by $\frac{\Delta}{2}eE_{c,z} = \lambda_{SO}$ and the material becomes a perfect semimetal. This critical electric field for germanene is 0.72 V/nm, i.e. only slightly larger than the electric field caused by the work function difference between germanene and W scanning tunneling microscopy tip. In order to be able to measure the spin-orbit gap in germanene with scanning tunneling spectroscopy, the work function of the scanning tunneling microscopy tip should be considerably lower than 4.5 eV. For instance a scanning tunneling microscopy tip work function of 4.0 eV would result in a 7 meV reduction of the spin-orbit gap. Unfortunately, all metals that can be used as scanning tunneling microscopy tips have work functions in the range of 4.5 eV to 5.5 eV. If the electric field exceeds the critical value $E_{c,z}$, the bandgap reopens and the germanene becomes a normal band insulator (see Figure 2). It should be noted here that the effect of the electric field on the size of the spin-orbit gap is independent on the sign of the electric field. As a final remark we would like to emphasize that the scanning tunneling spectroscopy measurements are performed at non-zero temperatures and therefore thermal broadening effects have to be taken into account as well. So, even if there is a small bandgap, it is masked by thermal broadening [15]. We have to conclude that scanning tunneling spectroscopy is not the ideal technique to probe the spin-orbit gap and the quantum spin Hall effect in two-dimensional Dirac materials, such as silicene and germanene. Alternatively, angle-resolved photoemission could be used to verify the existence of a spin-orbit gap in germanene. We would like to reiterate that the spin-orbit bandgap is only affected by an external electric field if the honeycomb lattice is buckled. The latter implies that the electronic band structure of graphene, which has a perfectly planar honeycomb lattice, is not affected by an external electric field. On the other hand, the spin-orbit gap of stanene, for example, is about 100 meV [19]. Of course the actual size of the spin-orbit gap might still be affected by the electric field in the scanning tunneling microscopy junction, but it is very unlikely that this field is large enough to close the spin-orbit gap completely.

**Conclusions**

The mystery regarding the absence of a spin-orbit gap in scanning tunneling spectroscopy measurements of germanene has been solved. We found that scanning tunneling spectroscopy is not the ideal technique to probe the spin-orbit gap and the quantum spin Hall effect in germanene owing to its very low work function of only 3.8 eV and buckled honeycomb structure. The difference in work function between germanene and scanning tunneling microscopy tip (for instance 4.5 eV for W and 5.5 eV for Pt/Ir) results in a substantial electric field of the order of 1 eV/nm that has a dramatic effect on the measured size of the spin-orbit gap of germanene. We anticipate that our findings are also relevant for other two-dimensional Dirac materials that have a buckled honeycomb structure.

**Acknowledgments**

CC and HJWZ acknowledge the *Nederlandse Organisatie voor Wetenschappelijk Onderzoek* (NWO) for financial support.